\documentclass[showpacs,amsmath,amssymb,floatfix,prl,twocolumn]{revtex4}
\usepackage{graphicx,amsmath}

\begin{document}


\title {Towards physical laws for software architecture}

\author{ A.D. Chepelianskii }
\affiliation{LPS, Univ. Paris-Sud, CNRS, UMR 8502, F-91405, Orsay, France }

\begin{abstract}
Starting from the pioneering works on software architecture precious
guidelines have emerged to indicate how computer programs should be organized.
For example the ``separation of concerns''
suggests to split a program into modules 
that overlap in functionality as little as possible.
However these recommendations are mainly conceptual and are thus hard to
express in a quantitative form.
Hence software architecture relies on the individual
experience and skill of the designers rather than on quantitative laws.
In this article I apply the methods developed for the
classification of information on the World-Wide-Web to study
the organization of Open Source programs in an attempt to
establish the statistical laws governing software architecture. 
\end{abstract}

\maketitle

The rapid increase in the size of software systems creates new challenges for the design and maintenance 
of computer software. Modern systems are constructed from many components forming 
a complex interdependent network. Starting from pioneering works in the early 1970s \cite{Knuth,Dijkstra,Parnas,Dijkstra2}
software architecture has developed in a mature field that provides precious guidelines for 
efficient software development \cite{Jacobson,Bass}. However these recommendations
are mainly conceptual and are thus difficult to express in a quantitative form.
In this article I construct the network formed by procedure calls in several open source programs
with emphasis on the code of the Linux kernel \cite{Linux}. 
The obtained networks have scale-free properties similar to hyperlinks on the World-Wide-Web
and other types of scale-free networks \cite{Watts,Barabasi,Newman,Dorogovtsev,Meyer}. 
Thus procedures can be ordered efficiently using the link analysis algorithms developed for web-pages \cite{PageRank,HITS}.
This allows to find automatically the important elements in the structure of a program and to propose
a quantitative criterion characterizing well organized software architectures. Finally I analyze the spectral properties 
of the transition matrix between the procedures and compare it with recent results for other networks \cite{Toulouse}.

\begin{figure}
\begin{center}
\includegraphics[clip=true,width=8cm]{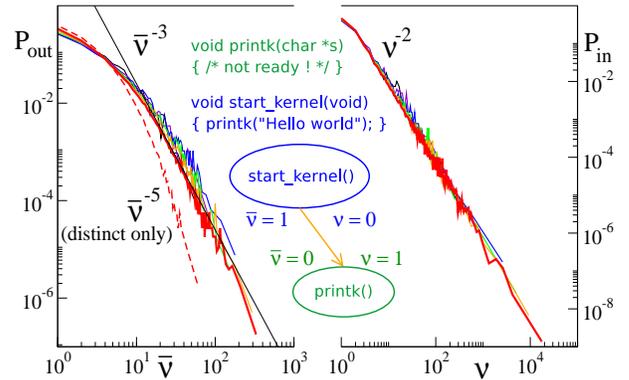}
\vglue -0.25cm
\caption{The diagram in the center represents the PCN of a toy kernel with two procedures 
written in the C programming language. The graph on the left/right shows the out/in degree probability 
distribution $P_{out}({\bar \nu})$/$P_{in}(\nu)$. The colors correspond to different Kernel releases.
The most recent version 2.6.32 with $N = 285509$ and an average $3.18$ calls per 
procedure is represented in red.
Older versions (2.4.37.6, 2.2.26, 2.0.40, 1.2.12, 1.0) with $N$ respectively equal to (85756, 38766, 14079, 4358, 2751) follow the same behavior. 
The dashed curve shows the out-degree probability distribution if only calls to distinct 
destination procedures are kept.
}
\label{LinuxFig1}
\end{center}
\end{figure}

In order to analyze quantitatively the network properties of computer code, 
I study several open source programs written in the C programming language \cite{LanguageC}.
In this widespread language the code is structured as a sequence 
of procedures calling each other, thus the organization of a program can be naturally 
represented as a procedure call network (PCN) where each node represents a 
procedure and each oriented edge corresponds to a procedure call.
This network is built by scanning lexically the source code of a project,
identifying all the defined procedures. For each of them a list keeps track 
of the procedures calls inside their definition.
An example of the obtained network for a toy code with two procedures is shown on  Fig.~\ref{LinuxFig1}.

The out/in-degrees of a node $i$ in this network are noted ${\bar \nu}(i)$ and $\nu(i)$ respectively.
The values of these numbers for the toy code are also given on Fig.~\ref{LinuxFig1},
they correspond to the number of out/in-going calls for each procedure.
A network is called scale-free, when the distributions of the degrees $\nu$ and ${\bar \nu}$ 
are characterized by power-law tails. Many networks in nature and in computer science fall in this class,
for example this is the case for the World-Wide-Web (WWW) \cite{Meyer,donata,Pandurangan,Baldi,Avrachenkov} 
and for the package dependencies in 
Linux distributions \cite{LinuxPackage}. 
The degree distributions for the PCN of several releases of the Linux kernel
are presented on Fig.~\ref{LinuxFig1}. They show unambiguously that PCN is a 
scale-free network with properties similar to WWW. Indeed the decay of the probability 
distribution $P_{in}(\nu)$ of in-going calls is well described 
by the power law $P_{in}(\nu) \propto \nu^{-\gamma_{in}}$ 
with $\gamma_{in} = 2.0 \pm 0.02$. 
The probability distribution of out-going calls also follows a power law
$P_{out}({\bar \nu}) \propto {\bar \nu}^{-\gamma_{out}}$ with $\gamma_{out} = 3.0 \pm 0.1$.
These values are close to the exponents found in the WWW where $\gamma_{in} = 2.1$ and 
$\gamma_{out} \approx 2.7$ \cite{donata,Pandurangan}. In the above distributions all procedure calls were 
included, if only calls to distinct functions are counted in the out-degree distribution 
the exponent drops to $\gamma_{out} \approx 5$ whereas $\gamma_{in}$ remains unchanged.
It should be stressed that the distributions for the different kernel releases remain stable 
even if the network size increases from $N = 2751$ for version 1.0 
to $N = 285509$ for the latest 2.6.32 version.

This similarity between PCN and WWW networks can be attributed to important 
development constraints that exist for both networks. Indeed WWW was designed 
as an information sharing system where users can easily access and create entries. 
The same principle applies also for Open Source development where the project 
is advanced by a loosely-knitted programmer community. 

Due to this similarity it is natural to apply the methods developed to 
organize information on the WWW to the PCN. PageRank is probably  the most successful
known link analysis algorithm \cite{PageRank}. It is based on the construction 
of the Google matrix :
\begin{align}
  G_{ij} = \alpha S_{ij} + (1-\alpha) / N 
\end{align} 
where the matrix $S$ is constructed by normalizing to unity all columns of the adjacency matrix,
and replacing columns with zero elements by $1/N$, $N$ being the network size \cite{Meyer}.
The damping parameter $\alpha$, in the WWW context describes the probability 
to jump to any node for a random surfer. 
For PCN this parameter can describe the 
probability to modify a global variable that affects the overall code behavior. 
The value $\alpha \approx 0.85$ seems to give 
a good classification \cite{Meyer} for WWW, thus I also used this value for PCN.
The matrix $G$ belongs to the class of Perron-Frobenius operators. Its largest eigenvalue 
is $\lambda = 1$ and other eigenvalues have $|\lambda| \le \alpha$. The right eigenvector 
at $\lambda = 1$ gives the probability $\rho(i)$ to find a random surfer at site $i$;
it is called the PageRank vector. Once the PageRank is found, 
WWW sites are sorted by decreasing $\rho(i)$, the site rank in this index $K(i)$
reflects the site relevance.

\begin{figure}
\begin{center}
\includegraphics[clip=true,width=8cm]{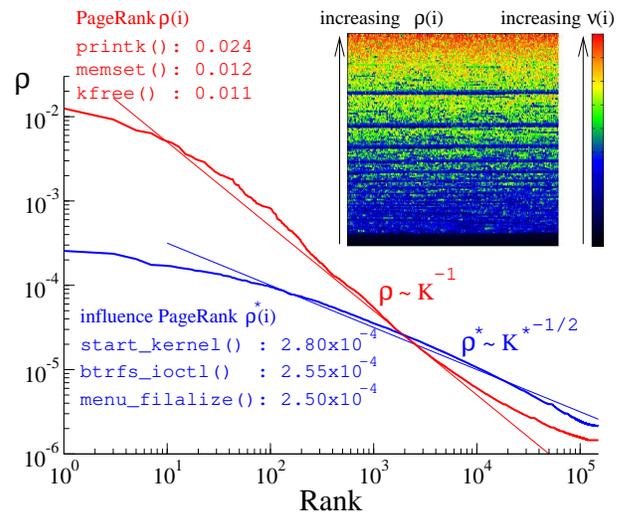}
\vglue -0.25cm
\caption{PageRank $\rho$ and influence-PageRank $\rho^*$ as a function of the 
ranks $K$ (for $\rho$) and $K^*$ (for $\rho^*$) for the PCN of the Linux Kernel, release 2.6.32.
The procedures with highest $\rho$ and $\rho^*$ are given on the left.
The inset illustrates the correlation between $\rho$ and the in-degree $\nu$:
procedures are serpentine ordered from low $\rho$ at the bottom to high $\rho$ on the top,
while the color code follows the value of the in-degree. 
}
\label{FigPageRank}
\end{center}
\end{figure}

The PageRank $\rho$ for the Linux PCN is shown on Fig.~\ref{FigPageRank} as a function of rank $K$.
The decay of $\rho(K)$ is well described by a power-law $\rho(K) \propto K^{-\beta}$ with 
$\beta \approx 1$, this value is consistent with the relation $\beta = 1/(\gamma_{in}-1)$ 
which would be exact if the PageRank of a procedure was proportional to its in-degree $\nu$.
It is known that for WWW this proportionality is qualitatively valid \cite{Pandurangan} although 
the PageRank classification introduces significant mixing compared to a classification 
based only on the in-degree distribution. The inset on Fig.~\ref{FigPageRank} illustrates
that this mixing exists also for PCN, hence PageRank classification for procedures 
is expected to be more informative and stable as in WWW. Fig.~\ref{FigPageRank} also 
reports the three procedures with the highest PageRank in the Linux Kernel. 
These popular procedures perform well defined tasks which may be useful in any part of the code: 
for example $printk()$ reports system messages and $memset(),\;kfree()$ intervene in memory allocation.

Although these procedures with high PageRank take care of highly useful tasks, 
their role in the overall program structure is limited. This suggests the existence of
 another complementary classification reflecting the procedure influence on the code organization. 
In the Hubs and Authority algorithm \cite{HITS} proposed in the WWW context,
the sites are characterized by two ranks reflecting their ``hubness'' (influence)
and ``authority'' (popularity). However this method 
is less stable than PageRank and is generally used for small subnetworks \cite{HITS,Baldi}. 
Hence I apply an alternative approach which is still based on the PageRank algorithm.
It consists in inverting the direction of links in the adjacency matrix before 
the construction of the Google matrix. This transposed adjacency matrix describes the 
flow of information returned from the called procedures to their parents. 
I will call influence-PageRank $\rho^*(i)$ the PageRank vector of this modified Google matrix,
the procedures can now be sorted according to their influence $\rho^*(i)$ yielding 
a new rank $K^*(i)$. 
The dependence of $\rho^*(i)$ on $K^*(i)$ for the Linux Kernel code is presented on Fig.~\ref{FigPageRank}.
Again the decay is well described by $\rho^*(K^*) \propto {K^*}^{-\beta^*}$ where 
$\beta^* \approx 1/(\gamma_{out}-1) \approx 1/2$. In this classification, the first 
procedures fulfill an important organizational role: e.g. $start\_kernel()$ 
initializes the Kernel and manages the repartition of tasks.

\begin{figure}
\begin{center}
\includegraphics[clip=true,width=8cm]{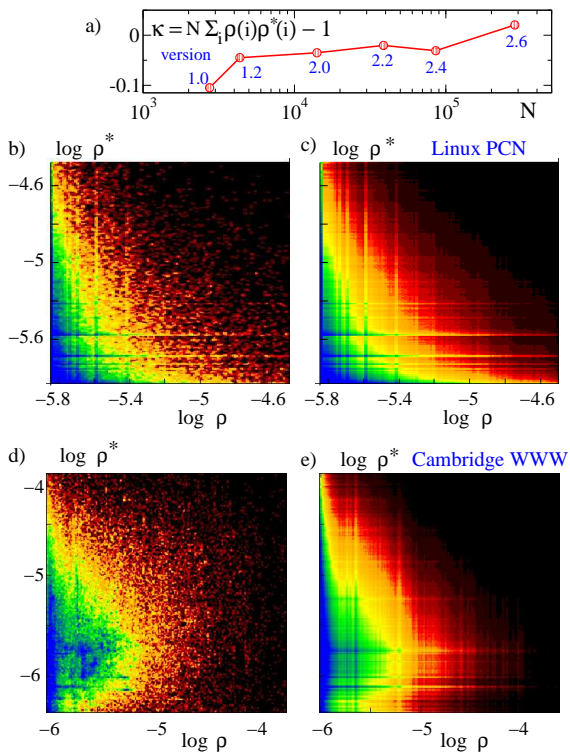}
\vglue -0.25cm
\caption{The left panel b) represents the joint 
probability distribution $P(\rho, \rho^*)$ as a function of $\log \rho$ and $\log \rho^*$ 
for the PCN of the Linux Kernel, release 2.6.32.
Regions with low probability are colored in Black/Red, while 
high probability are colored in Blue/Green.  The panel c) shows the
product probability $p(\rho) p^*(\rho^*)$ on the same scale,
it reproduces $P(\rho, \rho^*)$ with a high fidelity. 
The panel a) shows the value of the correlator $\kappa$ as a function of PCN size $N$ 
for the Linux Kernel releases from Fig.~\ref{LinuxFig1}. 
The two panels d) and e) compare the joint 
probability distribution $P(\rho, \rho^*)$ with the product probability $p(\rho) p^*(\rho^*)$ 
for the Cambrdige University WWW network.
The correlated structure along the diagonal $\rho = \rho^*$ which is present in panel d) is not reproduced on panel e). 
}
\label{FigCorrel}
\end{center}
\end{figure}

The correlation between popular and influential procedures in the PCN  
network is described by the joint probability distribution $P(\rho, \rho^*)$ 
that gives the probability of finding a procedure $i$ with $(\rho(i), \rho^*(i))$ 
in a small area around $(\rho, \rho^*)$. This distribution is displayed on 
Fig.~\ref{FigCorrel}  where it is compared with the distribution that
is obtained under the assumption that $\rho$ and $\rho^*$ are independent 
quantities. This distribution stems from the product of probabilities 
$p(\rho)$ and $p^*(\rho^*)$ to find a procedure in an interval around $\rho$ and $\rho^*$ respectively
so that $P = p(\rho) p^*(\rho^*)$. These two distributions are very similar, 
showing that the popularity and influence are weakly correlated in the PCN network.
The direct computation of the correlator $\kappa$ :
\begin{align}
\kappa = N \sum_i \rho(i) \rho^*(i) - 1
\end{align}
supports this assumption of independence. 
Indeed it was found that $|\kappa| \ll 1$ for the PCN of the Linux Kernel
for all releases. For most releases this correlator is negative 
indicating a certain anti-correlation between popular and influential procedures. 
These observation hold also for other OpenSource software 
including Gimp 2.6.8 ($\kappa=-0.068, N=17540$) and X Windows server R7.1-1.1.0 ($\kappa=-0.027, N=14887$).

This absence of correlations between popularity and influence in PCN contrasts 
with the WWW hyperlink network. In the latter case, the correlator is positive 
and of order unity: this was confirmed by analyzing hyperlinks for several UK 
universities available at \cite{UkUniv}. For example, I find for the web sites of Universities at 
Cambridge ($\kappa=3.79, N=376836$), Oxford ($\kappa=1.52,N=331955$), 
Bath ($\kappa=7.22, N=112143$) and Hull ($\kappa=2.09,N=21061$).
Note that the typical vale of $\kappa$ does not directly depend on the network size.
The joint probability $P(\rho, \rho^*)$ and the product probability $p(\rho) p^*(\rho^*)$ 
for the Cambrdige University network are compared on Fig.~\ref{FigCorrel}.
The product probability reproduces to some extent the behavior of $P(\rho, \rho^*)$ 
but fails to capture the correlations along the diagonal $\rho = \rho^*$
as expected from the positive value of the correlator $\kappa=3.79$.

The above observations suggest that the independence 
between popular procedures, fulfilling important but well defined 
tasks, and influential procedures, which organize and assign tasks 
in the code, is an important ingredient of well structured software.
The heuristic content of this independence criterion is 
related to the well-known concept of ``separation of concerns'' \cite{Dijkstra2} in software architecture. 
The correlation coefficient $\kappa$ allows to express this concept in a quantitative way.
Procedures that have high values of both $\rho(i)$ and $\rho^*(i)$ can therefore 
play a critical role since they are popular and influential at the same time.
For example in the Linux Kernel, $do\_fork()$ that creates new processes belongs to this class.
These critical procedures may introduce subtle errors because they entangle
otherwise independent segments of code.

\begin{figure}
\begin{center}
\includegraphics[clip=true,width=8cm]{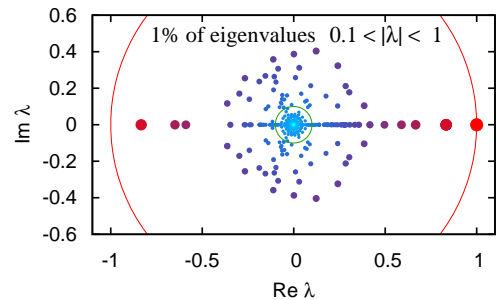}
\vglue -0.25cm
\caption{Distribution of eigenvalues $\lambda$ in the complex plane for the Google matrix 
of the Linux Kernel 2.0.40 with $N = 14079$. Circles highlight the ring region $0.1 < |\lambda| < 1$.
}
\label{FigEigenvalues}
\end{center}
\end{figure}

The eigenvalues of the matrix $G$ provide information on the relaxation 
rates to the PageRank. Eigenvalues with $|\lambda|$ close to unity,
represent independent component weakly connected with the rest of the network.
The WWW has a significant number of such modes \cite{Meyer,Avrachenkov,Toulouse}
showing the existence of many independent communities.
A typical eigenvalue distribution in the complex plane for Linux PCN is shown on Fig.~\ref{FigEigenvalues}.
The proportion of modes with $|\lambda| > 0.1$ is very small (around $1\%$ for network size $N = 14079$)
compared to the case of University networks \cite{Toulouse} where this percentage is around $50\%$ 
(for example for the Liverpool John Moores University with $N = 13578$).
This result can be interpreted as follows: the web contains many quasi-independent 
communities whereas the PCN must ensure a strong coordination between 
the different procedures that therefore must be able to exchange information.

The presented studies demonstrate close similarities between  
software architecture and scale-free networks especially 
with the World-Wide-Web. However they show that these networks 
have also substantial differences: the absence of correlation 
between popularity and influence in procedure call networks, 
and a large number of vanishing eigenvalues in the Google matrix
which indicates on the small number of independent communities in computer codes. 
The properties of software networks found here may lay the foundation 
for a quantitative description of functional software architectures.
The proposed methods can be generalized to object oriented programming 
and may find several applications in software development.
Possible applications include indications for the conception of code documentation
and improvements in code refactoring techniques. Finally the identification 
of critical procedures may facilitate the correction of subtle errors 
that arise due to unintended entanglement in the code.  

I thank T.C. Phan for fruitful discussions on software development 
and acknowledge DGA for support.


\begin{thebibliography}{99}

\bibitem{Knuth} Donald E. Knuth {\it The Art of Computer Programming}, Addison-Wesley, ISBN 0-201-03801-3 (1968)

\bibitem{Dijkstra} E.W. Dijkstra, {\it Go to statement considered harmful}, Comm. ACM, {\bf 11}, 147 (1968)

\bibitem{Parnas} D.L. Parnas, {\it On the criteria to be used in decomposing systems into modules}, Comm. ACM, {\bf 15}, 1053 (1972)

\bibitem{Dijkstra2} E.W. Dijkstra, {\it Selected writings on Computing: A Personal Perspective}, Springer-Verlag New York, ISBN 0-387-90652-5 (1982)

\bibitem{Jacobson} I. Jacobson {\it Object-oriented software engineering}, 
Addison-Wesley (1992) ISBN 0-201-54435-0 

\bibitem{Bass} L. Bass, P. Clements, R. Kazman {\it Software Architecture in Practice}, 
Addison-Wesley (2003) ISBN-10: 0-321-15495-9

\bibitem{Linux} The source code of the different Linux Kernel releases were downloaded from http://www.kernel.org/

\bibitem{Watts} D.J.Watts and S.H.Strogatz, {\it Collective dynamics of ``small-world'' networks (1998)}, Nature {\bf 393}, 440 (1998)

\bibitem{Newman} M. E. J. Newman, {\it The structure of scientific collaboration networks}, 
Proc. Natl. Acad. Sci. USA {\bf 98}, 404 (2001). 

\bibitem{Barabasi} R. Albert, A.-L. Barabási, {\it Statistical mechanics of complex networks}, Rev. Mod. Phys. {\bf 74}, 47(2002). 

\bibitem{Dorogovtsev} S.~N.~Dorogovtsev and J.~F.~F.~Mendes, {\it Evolution of 
                     Networks}, Oxford University Press (Oxford, 2003).

\bibitem{Meyer} A.~M.~Langville and C.~D.~Meyer, {\it Google's PageRank
  and Beyond: The Science of Search Engine Rankings}, Princeton University
  Press (Princeton, 2006)


\bibitem{PageRank} S.~Brin and L.~Page, {\it The anatomy of a largescale hypertextual web search engine}, 
Computer Networks and ISDN Systems {\bf 30}, 107 (1998).

\bibitem{HITS} J. Kleinberg , {\it Authoritative sources in a hyperlinked environment}, Jour. ACM {\bf 46}, 604 (1999)

\bibitem{Toulouse} O.~Giraud, B.~Georgeot and D.~L.~Shepelyansky, {\it "Delocalization transition for the Google matrix}
    Phys. Rev. E {\bf 80}, 026107 (2009); B. Georgeot, O. Giraud, D.L. Shepelyansky {\it"Spectral properties of the Google matrix of the World Wide Web and other directed networks} arXiv:1002.3342 (2010)

\bibitem{LanguageC} B.W. Kernighan and D.M. Ritchie {\it The C Programming Language} Englewood Cliffs, NJ: Prentice Hall. ISBN 0-13-110163-3.

\bibitem{donata} D.~Donato, L.~Laura, S.~Leonardi and 
              S.~Millozzi, {\it  Large scale properties of the Webgraph}  Eur. Phys. J. B {\bf 38}, 239 (2004)

\bibitem{Pandurangan} G.~Pandurangan, P.~Raghavan and E.~Upfal, {\it  Using PageRank to Characterize Web Structure}, Internet Math. {\bf 3}, 1 (2005).

\bibitem{Baldi} P. Balid, P. Frasconi and P. Smyth {\it Modeling the Internet and the Web :  Probabilistic Methods and Algorithms},
        Published by John Wiley \& Sons, Ltd. ISBN: 0-470-84906-1 (2003)

\bibitem{Avrachenkov} K.~Avrachenkov, D.~Donato and N.~Litvak (Eds.),
        {\it Algorithms and Models for the Web-Graph: 
        6th International Workshop, 
        WAW 2009 Barcelona,  Proceedings},  Springer-Verlag, Berlin,
        Lecture Notes Computer Sci. {\bf 5427}, Springer, Berlin (2009).


\bibitem{LinuxPackage} T. Maillart, D. Sornette, S. Spaeth and G. von Krogh  {\it Empirical Tests of Zipf's Law Mechanism in Open Source Linux Distribution}, 
       Phys. Rev. Lett. {\bf 101}, 218701 (2008) 

\bibitem{UkUniv} Academic Web Link Database Project http://cybermetrics.wlv.ac.uk/database/



\end{thebibliography}
\end{document}